\documentclass[prd,aps,singlecolumn,preprintnumbers, showpacs, nofootinbib,superscriptaddress,notitlepage]{revtex4-1}
\usepackage{amssymb, amsthm}
\usepackage{amsmath}    
\usepackage{mathrsfs}   
\usepackage{graphicx}   
\usepackage{color}      
\usepackage{footnote}   

\usepackage{slashed}    
\usepackage{subfigure}  
\usepackage{hyperref}   

\usepackage{rotating}   
\usepackage{multirow}   
\usepackage[normalem]{ulem} 
\usepackage[utf8]{inputenc}

\newcommand{\beq}{\begin{eqnarray}}
\newcommand{\eeq}{\end{eqnarray}}

\newcommand{\Blue}[1]{\textcolor{black}{#1}}

\begin{document}

\preprint{MSUHEP-17-015, MIT-CTP/4942}

\title{Symmetry Properties of Nonlocal Quark Bilinear Operators on a Lattice}

\collaboration{\bf{$\rm {\bf LP^3}$ Collaboration}}

\author{Jiunn-Wei Chen}
\email{jwc@phys.ntu.edu.tw}
\affiliation{Department of Physics, Center for Theoretical Physics, and Leung Center for Cosmology and Particle Astrophysics, National Taiwan University, Taipei 106, Taiwan}
\affiliation{Center for Theoretical Physics, Massachusetts Institute of Technology, Cambridge, MA 02139, USA}

\author{Tomomi Ishikawa}
\email{tomomi.ik@gmail.com}
\affiliation{T.~D.~Lee Institute, Shanghai Jiao Tong University,
Shanghai, 200240, P. R. China}

\author{Luchang Jin}
\affiliation{Physics Department, University of Connecticut,
Storrs, Connecticut 06269-3046, USA}
\affiliation{RIKEN BNL Research Center, Brookhaven National Laboratory,
Upton, NY 11973, USA}

\author{Huey-Wen Lin}
\affiliation{Department of Physics and Astronomy, Michigan State University, East Lansing, MI 48824, USA}
\affiliation{Department of Computational Mathematics, Michigan State University, East Lansing, MI 48824, USA}


\author{Jian-Hui Zhang}
\email{jianhui.zhang@ur.de}
\affiliation{Institut f\"ur Theoretische Physik, Universit\"at Regensburg, D-93040 Regensburg, Germany}

\author{Yong Zhao}
\affiliation{Center for Theoretical Physics, Massachusetts Institute of Technology, Cambridge, MA 02139, USA}


\begin{abstract}
Using symmetry properties, we determine the mixing pattern of a class of nonlocal quark bilinear operators containing a straight Wilson line along a spatial direction.  
We confirm the previous study that mixing among the lowest dimensional operators, which have mass dimension equals three, can occur if chiral symmetry is broken in the lattice action. For higher dimensional operators, we find that the dimension three 
operators will always mix with dimension four 
operators even if chiral symmetry is preserved. Also, the number of 
 dimension four
operators involved in the mixing is large hence it is impractical to remove the mixing by the improvement procedure. Our result is important to determining the Bjorken-$x$ dependence parton distribution functions using the quasi-distribution method on a Euclidean lattice. The requirement of using large hadron momentum in this approach makes the control of 
errors from dimension four operators
even more important.
\end{abstract}

\maketitle

\section{Introduction}
\label{SEC:Introduction}

Taming systematic uncertainties is critical to obtain meaningful results in lattice QCD. For example, the nonperturbative renormalization method of the Rome-Southampton collaboration~\cite{Martinelli:1994ty} has been widely used to convert from the lattice scheme to continuum schemes, avoiding the introduction of errors from slowly converging lattice perturbation theory. Another example is use of Symanzik improvement~\cite{Symanzik:1983dc,Luscher:1984xn} to systematically reduce discretization errors due to nonzero lattice spacing $a$. Since it is crucial to understand the mixing patterns of the operators involved, understanding the symmetries of a problem provides a powerful nonperturbative method. Symmetries could protect certain mixings from happening, while those not protected by symmetries could happen under quantum corrections. Although symmetry considerations do not provide a quantitative analysis of the mixing, they do provide a complete mixing pattern among operators in the problem.

In this work, we use the symmetries of lattice QCD to analyze the mixing pattern for a class of nonlocal quark bilinear operators defined in Eq.~\eqref{NL}.
Their renormalization in the continuum has been discussed since the 1980s~\cite{Craigie:1980qs, Dorn:1986dt}. In recent years, there has been renewed interest in the renormalization of these operators in the context methods for calculating the Bjorken-$x$ dependence of the hadron parton distribution functions (PDFs) using lattice QCD: the quasi-PDF method~\cite{Ji:2013dva} and its variations~\cite{Ma:2014jla, Radyushkin:2017cyf}. For recent progress in this area, see Refs.~\cite{Lin:2014zya,Chen:2016utp,Lin:2017ani,Alexandrou:2015rja,Alexandrou:2016jqi,Alexandrou:2017huk,Chen:2017mzz,
Zhang:2017bzy,Chen:2017gck,Xiong:2013bka,Ji:2015jwa,Ji:2015qla,Xiong:2015nua,Ji:2014hxa,Monahan:2017hpu,
Ji:2018hvs,Stewart:2017tvs,Constantinou:2017sej,Green:2017xeu,Izubuchi:2018srq,Xiong:2017jtn,Wang:2017qyg,
Wang:2017eel,Xu:2018mpf,Ishikawa:2016znu,Chen:2016fxx,Ji:2017oey,Ishikawa:2017faj,Chen:2017mie,
Chen:2017lnm,Li:2016amo,Monahan:2016bvm,Radyushkin:2016hsy,Rossi:2017muf,Carlson:2017gpk,Ji:2017rah,
Chen:2018xof,Alexandrou:2018pbm,Gamberg:2014zwa,Nam:2017gzm,Broniowski:2017wbr,Jia:2017uul,
Hobbs:2017xtq,Jia:2015pxx,Chen:2018fwa,Orginos:2017kos,Radyushkin:2018cvn,Zhang:2018ggy}. 
A special feature of these nonlocal quark bilinears is that the Wilson line connecting the quark fields receives power-divergent contributions. 
A nonperturbative subtraction of the power divergence was proposed in Refs.~\cite{Musch:2010ka, Ishikawa:2016znu, Chen:2016fxx} by recasting the Wilson line as a heavy-quark field in the auxiliary-field approach~\cite{Craigie:1980qs, Dorn:1986dt} such that the counterterm needed to subtract the power divergence is just the counterterm for heavy-quark mass renormalization. The renormalization for the nonlocal quark bilinears in the continuum was studied
in Refs.~\cite{Ji:2015jwa,Ji:2017oey, Ishikawa:2017faj} and on a lattice in Ref.~\cite{Green:2017xeu}, and in nonperturbative renormalization schemes~\cite{Alexandrou:2017huk,Chen:2017mzz}.


A lattice theory has fewer symmetries than its corresponding continuum theory. This implies that there will be more mixing among operators in a lattice theory than in the corresponding continuum theory. For example, a pioneering one-loop lattice perturbation theory calculation using Wilson fermions showed that the breaking of chiral symmetry for the Wilson fermions induces the mixing shown in Eq.~\eqref{3p8}~\cite{Constantinou:2017sej}. In this work, instead of performing explicit computations, we use symmetries to systematically study the mixing patterns among nonlocal quark bilinears (part of this work was reported in \cite{Chen:2017mzz,Ishikawa:2018wun}). We study not only the mixing among the lowest dimensional nonlocal quark bilinears of mass dimension three
as Ref.~\cite{Constantinou:2017sej} did, but also the mixing between dimension three and dimension four operators,
which cannot be avoided even if chiral symmetry is preserved.\footnote{This is different for the case of local operators, where mixing between the dimension three and four
operators is forbidden by symmetries. If a lattice action is
${\cal O}(a)$-improved, then the mixing of dimension three and four
{\it local} operators is forbidden, but the mixing between {\it nonlocal} operators is still allowed.}
This feature is confirmed by the computation of an example one-loop diagram.

Our study is particularly relevant to the quasi-PDF approach, which receives power corrections in inverse powers of hadron momentum. It is important to find the window where hadron momentum is large enough to suppress power corrections (good progress was made using momentum smearing~\cite{Bali:2016lva,Alexandrou:2016jqi}), but small enough that mixing with dimension four operators
is under control. 
In the following, we first review the symmetry analysis of local quark bilinear operators, and then move to the nonlocal ones.

\section{Review of local quark bilinear operators}
\label{SEC:local}

If the $\theta$ term is neglected, the lattice action exhibits important discrete symmetries: the action is invariant under discrete parity (${\cal P}$), time reversal (${\cal T}$) and charge conjugation (${\cal C}$) transformations (see e.g. Ref.\cite{Gattringer:2010zz}). Chiral symmetry, which is a continuous symmetry, however, might be broken after the fermion fields are discretized. In this section, we review the symmetry properties for a specific set of local quark bilinear operators under these transformations. Then, we extend the analysis to nonlocal quark bilinear operators in the next section. The importance of these analyses is that if two operators transform differently, then symmetries will protect them from mixing with each other under quantum corrections to all orders in the coupling. Operators not protected from mixing by symmetries, in general, will mix.


\subsection{${\cal P}$, ${\cal T}$, ${\cal C}$ and axial transformations}

In this subsection, we summarize the transformations of fields under ${\cal P}$, ${\cal T}$, ${\cal C}$ and the axial transformation (the vector transformation in chiral symmetry is conserved in all the operators that we study). We work in Euclidean spacetime with coordinates $(x, y, z, \tau) = (1, 2, 3, 4)$ throughout this paper.  Gamma matrices are chosen to be Hermitian: $\gamma_{\mu}^{\dagger}=\gamma_{\mu}$, and $\gamma_5=\gamma_1\gamma_2\gamma_3\gamma_4$.

Since there is no distinction between time and space in Euclidean space, the parity transformation, denoted ${\cal P}_{\mu}$ with $\mu\in\{1,2,3,4\}$, can be defined with respect to any direction.
\begin{eqnarray}
\psi(x)&\xrightarrow[]{{\cal P}_{\mu}}&
\psi(x)^{{\cal P}_{\mu}}=\gamma_{\mu}\psi(\mathbb{P}_{\mu}(x)),\\
\overline{\psi}(x)&\xrightarrow[]{{\cal P}_{\mu}}&
\overline{\psi}(x)^{{\cal P}_{\mu}}=\overline{\psi}(\mathbb{P}_{\mu}(x))\gamma_{\mu},\\
U_{\nu\not=\mu}(x)&\xrightarrow[]{{\cal P}_{\mu}}&
U_{\nu\not=\mu}(x)^{{\cal P}_{\mu}}
=U_{\nu\not=\mu}^{\dagger}(\mathbb{P}_{\mu}(x)-\hat{\nu}),\\
U_{\mu}(x)&\xrightarrow[]{{\cal P}_{\mu}}&
U_{\mu}(x)^{{\cal P}_{\mu}}=U_{\mu}(\mathbb{P}_{\mu}(x)),
\end{eqnarray}
where $\mathbb{P}_{\mu}(x)$ is the vector $x$ with sign flipped except for the $\mu$-direction.

Analogously, time reversal transformation, denoted as ${\cal T}_{\mu}$, can be generalized in any direction in Euclidean space.
 \begin{eqnarray}
\psi(x)&\xrightarrow[]{{\cal T}_{\mu}}&
\psi(x)^{{\cal T}_{\mu}}=\gamma_{\mu}\gamma_5\psi(\mathbb{T}_{\mu}(x)),\\
\overline{\psi}(x)&\xrightarrow[]{{\cal T}_{\mu}}&
\overline{\psi}(x)^{{\cal T}_{\mu}}=
\overline{\psi}(\mathbb{T}_{\mu}(x))\gamma_5\gamma_{\mu},\\
U_{\mu}(x)&\xrightarrow[]{{\cal T}_{\mu}}&
U_{\mu}(x)^{{\cal T}_{\mu}}=U_{\mu}^{\dagger}(\mathbb{T}_{\mu}(x)-\hat{\mu}),\\
U_{\nu\not=\mu}(x)&\xrightarrow[]{{\cal T}_{\mu}}&
U_{\nu\not=\mu}(x)^{{\cal T}_{\mu}}
=U_{\nu\not=\mu}(\mathbb{T}_{\mu}(x)),
\end{eqnarray}
where $\mathbb{T}_{\mu}(x)$ is the vector $x$ with sign flipped in the $\mu$-direction.

Charge conjugation ${\cal C}$ transforms particles into antiparticles,
\begin{eqnarray}
\psi(x)&\xrightarrow[]{\cal C}&
\psi(x)^{\cal C}=C^{-1}\overline{\psi}(x)^{\top},\\
\overline{\psi}(x)&\xrightarrow[]{\cal C}&
\overline{\psi}(x)^{\cal C}=-\psi(x)^{\top}C,\\
U_{\mu}(x)&\xrightarrow[]{\cal C}&
U_{\mu}(x)^{\cal C}=U_{\mu}(x)^{\ast}=(U_{\mu}^{\dagger}(x))^{\top},
\end{eqnarray}
and 
\begin{eqnarray}
C\gamma_{\mu}C^{-1}=-\gamma_{\mu}^{\top}, \qquad
C\gamma_5C^{-1}=\gamma_5^{\top}.
\end{eqnarray}

The continuous axial rotation ($\cal A$) of the fermion fields is
\begin{eqnarray}
\psi(x)\xrightarrow[]{\cal{A}}\psi'(x)=e^{i\alpha\gamma_5}\psi(x),\qquad
\overline{\psi}(x)\xrightarrow[]{\cal{A}}\overline{\psi}'(x)
=\overline{\psi}(x)e^{i\alpha\gamma_5},
\label{EQ:chiral_transformation}
\end{eqnarray}
where $\alpha$ is the $x$-independent rotation angle for the global transformation.\footnote{The anomaly induced by the single-flavor axial rotation is identical for all the operators that we study. Hence, it can be safely neglected in the operator classification.} 
The explicit axial symmetry breaking pattern induced by the quark mass $m$ can be studied by introducing a spurious transformation
\begin{eqnarray}
\label{axial}
m\xrightarrow[]{\cal{A}}e^{-i\alpha\gamma_5}me^{-i\alpha\gamma_5},
\end{eqnarray}
so that the quark mass term is invariant under this extended axial transformation.

\subsection{Dimension three local operators}

Now we study the transformation properties for a class of local quark bilinear operators of the form
\begin{eqnarray}
O_{\Gamma}=\overline{\psi}(x)\Gamma\psi(x),
\end{eqnarray}
with 
\begin{eqnarray}
\Gamma\in\{\bf{1}, ~\gamma_{\mu}, ~\gamma_5,
 ~i\gamma_{\mu}\gamma_5, ~\sigma_{\mu\nu}\},
\end{eqnarray}
where $\sigma_{\mu\nu}=\frac{i}{2}[\gamma_{\mu},\gamma_{\nu}]$. 
\Blue{Quantum loop effects for these operators are in powers of $\log a$.}
The Hermitian conjugate is
\begin{eqnarray}
(O_{\Gamma})^{\dagger} 
=-O_{\gamma_4\Gamma\gamma_4}
=-G_4(\Gamma)O_{\Gamma},
\end{eqnarray}
where $G_{\mu}(\Gamma)$, which has a value of either $+1$ or $-1$, satisfies
\begin{eqnarray}
\label{Gmu}
\gamma_{\mu}\Gamma\gamma_{\mu}=G_{\mu}(\Gamma)\Gamma.
\end{eqnarray}
Therefore, depending on $\Gamma$, the expectation value of $O_{\Gamma}$ can be purely real or imaginary.

\begin{table}[t]
\begin{center}
\small
\begin{tabular}{cccccc}
\hline\hline
& $\Gamma=\bf{1}
$ & $\gamma_{\mu}$ & $\gamma_5$ & $i\gamma_{\mu}\gamma_5$ &
$\sigma_{\mu\nu}$\\
\hline
${\cal P}_{\rho=\mu}$ & E &  E & O & O & O \\
${\cal P}_{\rho\not=\mu}$ &
E  & O & O  & E  & O${}_{(\rho=\nu)}$/E${}_{(\rho\not=\nu)}$ \\
${\cal T}_{\rho=\mu}$ & E &  O & O & E & O \\
${\cal T}_{\rho\not=\mu}$ &
E  & E & O  & O  & O${}_{(\rho=\nu)}$/E${}_{(\rho\not=\nu)}$ \\
${\cal C}$ & E & O & E & E & O \\
${\cal A}$ & V & I & V & I & V \\
\hline\hline
\end{tabular}
\caption{Properties of the dimension three
local operator $O_{\Gamma}$ under parity (${\cal P}_{\rho}$), time reversal (${\cal T}_{\rho}$), charge conjugation (${\cal C}$) and axial (${\cal A}$) transformations. E and O stand for even and odd while I and V stand for invariant and variant under transformations.}
\label{TAB:PTC_local_eo}
\end{center}
\end{table}

Under ${\cal P}$, ${\cal T}$, and ${\cal C}$, the local quark bilinear transforms as
\begin{eqnarray}
O_{\Gamma}\xrightarrow[]{{\cal P}_{\mu}}
O_{\gamma_{\mu}\Gamma\gamma_{\mu}},\;\;\;
O_{\Gamma}\xrightarrow[]{{\cal T}_{\mu}}
O_{\gamma_5\gamma_{\mu}\Gamma\gamma_{\mu}\gamma_5},\;\;\;
O_{\Gamma}\xrightarrow[]{\cal C}
O_{(C\Gamma C^{-1})^{\top}}.
\end{eqnarray}
$O_{\Gamma}$ either stays invariant (even, E) or changes sign (odd, O) under a transformation. The result is summarized in Table~\ref{TAB:PTC_local_eo}. Operators of different $\Gamma$ do not mix under renormalization, since they transform differently under ${\cal P}_{\mu}$ or ${\cal T}_{\mu}$. ${\cal C}$ alone does not protect the operators from mixing with each other. 

Under an axial rotation (with Eq.~\eqref{axial} included), $O_{\Gamma}$ is either invariant (I) or variant (V) as shown in Table~\ref{TAB:PTC_local_eo}. Some lattice fermions, such as Wilson fermions, break the axial symmetry, but from the above discussion, we see that axial symmetry is not essential in protecting $O_{\Gamma}$ from mixing. Only ${\cal P}_{\mu}$ or ${\cal T}_{\mu}$ is needed. 

\subsection{Dimension four local operators}

At dimension four, we can further classify the operators into $p$ type and $m$ type operators, which have one more insertion of derivative or quark mass, respectively, compared with the dimension three operators. Here $p$ denotes a typical momentum in the external state. It is useful to define covariant derivatives, $\overrightarrow{D}_{\mu}$ and $\overleftarrow{D}_{\mu}$. Acting on a field $\phi(x)$, 
\begin{eqnarray}
\overrightarrow{D}_{\mu}\phi(x)&=& 
\frac{1}{2a}\left[U_{\mu}(x)\phi(x+\hat{\mu}a)
-U_{\mu}^{\dagger}(x-\hat{\mu}a)\phi(x-\hat{\mu}a)\right],
\\
\phi(x)\overleftarrow{D}_{\mu}&=& 
\frac{1}{2a}\left[\phi(x+\hat{\mu}a)U_{\mu}^{\dagger}(x)
-\phi(x-\hat{\mu}a)U_{\mu}(x-\hat{\mu}a)\right].
\end{eqnarray}
Euclidean four-dimensional rotational symmetry dictates that $p$ type operators are constructed by inserting $\overrightarrow{\slashed{D}}$ and $\overleftarrow{\slashed{D}}$ into $O_{\Gamma}$:
\begin{eqnarray}
Q_{\Gamma\!\overrightarrow{D}}=
\overline{\psi}(x)\Gamma\overrightarrow{\slashed{D}}\psi(x),\qquad
&&
Q_{\overleftarrow{D}\!\Gamma}=
\overline{\psi}(x)\overleftarrow{\slashed{D}}\Gamma\psi(x),\\
Q_{\overrightarrow{D}\!\Gamma}=
\overline{\psi}(x)\overrightarrow{\slashed{D}}\Gamma\psi(x),\qquad
&&
Q_{\Gamma\!\overleftarrow{D}}=
\overline{\psi}(x)\Gamma\overleftarrow{\slashed{D}}\psi(x).
\end{eqnarray}
It can be shown that those operators transform in the same way as $O_{\Gamma}$ under ${\cal P}_{\mu}$ and ${\cal T}_{\mu}$, while under ${\cal C}$,  
\begin{eqnarray}
Q_{\Gamma\!\overrightarrow{D}/\overrightarrow{D}\!\Gamma}
\xrightarrow[]{\cal C}
-Q_{\overleftarrow{D}\!(C\Gamma C^{-1})^{\top}
/(C\Gamma C^{-1})^{\top}\!\overleftarrow{D}},\qquad
Q_{\overleftarrow{D}\!\Gamma/\Gamma\!\overleftarrow{D}}
\xrightarrow[]{\cal C}
-Q_{(C\Gamma C^{-1})^{\top}\!\overrightarrow{D}
/\overrightarrow{D}\!(C\Gamma C^{-1})^{\top}},
\end{eqnarray}
with $\overrightarrow{\slashed{D}}$ and $\overleftarrow{\slashed{D}}$ operators transforming into each other. 
Therefore, it is convenient to define the combinations
\begin{eqnarray}
O_{\Gamma}^{p(\pm)}=
Q_{\overleftarrow{D}\!\Gamma}\pm Q_{\Gamma\!\overrightarrow{D}},\qquad
O_{\overline{\Gamma}}^{p(\pm)}=
Q_{\Gamma\!\overleftarrow{D}}\pm Q_{\overrightarrow{D}\!\Gamma},
\label{EQ:Opa-operator}
\end{eqnarray}
which are either even or odd under  ${\cal C}$. The transformation properties of the $p$ type operators are listed in Table~\ref{TAB:PTC_local_Opa_eo}. By comparing with Table~\ref{TAB:PTC_local_eo}, we observe that ${\cal P}_{\mu}$, ${\cal T}_{\mu}$ and ${\cal C}$ symmetries do not protect $O_{\Gamma}$ from mixing with $O_{\Gamma/\overline{\Gamma}}^{p(-)}$, but axial symmetry does. So, if the lattice theory preserves axial or chiral symmetry, then the dimension three and $p$ type dimension four 
operators studied above will not mix.  

\begin{table}[t]
\begin{center}
\small
\begin{tabular}{cccccc}
\hline\hline
& $\Gamma=\bf{1}$ & $\gamma_{\mu}$ & $\gamma_5$ & $i\gamma_{\mu}\gamma_5$ &
$\sigma_{\mu\nu}$\\
\hline
${\cal P}_{\rho=\mu}$ & E &  E & O & O & O \\
${\cal P}_{\rho\not=\mu}$ &
E  & O & O  & E  & O${}_{(\rho=\nu)}$/E${}_{(\rho\not=\nu)}$ \\
${\cal T}_{\rho=\mu}$ & E &  O & O & E & O \\
${\cal T}_{\rho\not=\mu}$ &
E  & E & O  & O  & O${}_{(\rho=\nu)}$/E${}_{(\rho\not=\nu)}$ \\
${\cal C}(O_{\Gamma/\overline{\Gamma}}^{p(+)})$ & O & E & O & O & E \\
${\cal C}(O_{\Gamma/\overline{\Gamma}}^{p(-)})$ & E & O & E & E & O \\
${\cal A}$ & I & V & I & V & I \\
\hline\hline
\end{tabular}
\caption{Transformation properties of the dimension four $p$ type 
local operators  $O_{\Gamma/\overline{\Gamma}}^{p(\pm)}$. 
Notations are the same as in Table~\ref{TAB:PTC_local_eo}.}
\label{TAB:PTC_local_Opa_eo}
\end{center}
\end{table}

Now we consider $m$ type
operators. The only operator appears at this order is 
\begin{eqnarray}
O_{\Gamma}^{m}=m\overline{\psi}(x)\Gamma\psi(x),
\label{EQ:local_Oma}  
\end{eqnarray}
which transforms in the same way as $O_{\Gamma}$ under ${\cal P}_{\mu}$, ${\cal T}_{\mu}$ and ${\cal C}$. However, it transforms differently from $O_{\Gamma}$ under ${\cal A}$. 

Therefore, we conclude that if the lattice theory preserves axial or chiral symmetry, then the dimension three and dimension four
operators (including both the  
$p$ type and $m$ type operators) studied above will not mix.

\section{Nonlocal quark bilinear operators}
\label{SEC:non-local}

Having reviewed the operator-mixing properties of the local quark bilinears, we now apply the analysis to a specific type of nonlocal quark bilinears.

\subsection{Dimension three nonlocal operators}

We are interested in the nonlocal quark bilinear operators with quark fields separated by $\delta z$ in the $z$-direction:
\begin{eqnarray}
\label{NL}
O_{\Gamma}(\delta z)=
\overline{\psi}(x+\delta z)\Gamma U_3(x+\delta z;x)\psi(x),
\end{eqnarray}
where a straight Wilson line $U_3$ is added such that the operators are gauge invariant. Treating the $z$-direction differently from the other directions, we write
\begin{eqnarray}
\Gamma\in\{\bf{1}, ~\gamma_i, ~\gamma_3, ~\gamma_5, ~i\gamma_i\gamma_5,
~i\gamma_3\gamma_5, ~\sigma_{i3}, ~\epsilon_{ijk}\sigma_{jk}\},
\end{eqnarray}
where $i, j, k\not=3$. 
\Blue{These operators receive quantum loop corrections as powers of $1/a$ and $\log a$~\cite{Ji:2015jwa,Ji:2017oey, Ishikawa:2017faj}. It is important to keep in mind that one cannot take the continuum limit of the matrix elements of these operators.}

Under ${\cal P}_{\mu}$ and ${\cal T}_{\mu}$,
\begin{eqnarray}
O_{\Gamma}(\delta z)\xrightarrow[]{{\cal P}_{l\not=3}}
O_{\gamma_l\Gamma\gamma_l}(-\delta z),\qquad&&
O_{\Gamma}(\delta z)\xrightarrow[]{{\cal P}_3}
O_{\gamma_3\Gamma\gamma_3}(\delta z),
\label{EQ:parity-Pj-P3}\\
O_{\Gamma}(\delta z)\xrightarrow[]{{\cal T}_{l\not=3}}
O_{\gamma_5\gamma_l\Gamma\gamma_l\gamma_5}(\delta z),\qquad&&
O_{\Gamma}(\delta z)\xrightarrow[]{{\cal T}_3}
O_{\gamma_5\gamma_3\Gamma\gamma_3\gamma_5}(-\delta z).
\label{EQ:time-reversal-Tj-T3}
\end{eqnarray}
The transformations could change the sign of $\delta z$, so it is convenient to define
\begin{eqnarray}
\label{3p5}
O_{\Gamma\pm}(\delta z)=
\frac{1}{2}\left(O_{\Gamma}(\delta z)\pm O_{\Gamma}(-\delta z)\right),
\end{eqnarray}
whose Hermitian conjugate yields
\begin{eqnarray}
(O_{\Gamma\pm}(\delta z))^{\dagger} 
=\mp G_4(\Gamma)O_{\Gamma\pm}(\delta z).
\end{eqnarray}
Thus, the expectation value of $O_{\Gamma\pm}(\delta z)$ is either purely real or purely imaginary, depending on $\Gamma$. Under $\cal C$, 
\begin{eqnarray}
O_{\Gamma\pm}(\delta z)
\xrightarrow[]{\cal C}\pm O_{(C\Gamma C^{-1})^{\top}\pm}(\delta z).
\end{eqnarray}
\begin{table}[t]
\begin{center}
\small
\begin{tabular}{ccccccccc}
\hline\hline
& $\Gamma=\bf{1}_{+/-}$ & $\gamma_{i+/-}$ & $\gamma_{3+/-}$ & $\gamma_{5+/-}$ & $i\gamma_i\gamma_{5+/-}$ & $i\gamma_3\gamma_{5+/-}$ & $\sigma_{i3+/-}$ & $\epsilon_{ijk}\sigma_{jk+/-}$\\
\hline
${\cal P}_3$ & E & O & E & O & E & O & O & E\\       
${\cal P}_{l\not=3}$ & E/O & E/O$_{(l=i)}$ & O/E & O/E & O/E$_{(l=i)}$ & E/O & O/E$_{(l=i)}$ & E/O$_{(l=i)}$\\
&   & O/E$_{(l\not=i)}$ & & & E/O$_{(l\not=i)}$ & & E/O$_{(l\not=i)}$ & O/E$_{(l\not=i)}$\\
${\cal T}_3$ & E/O & E/O & O/E & O/E & O/E & E/O & O/E & E/O\\
${\cal T}_{l\not=3}$ & E & O$_{(l=i)}$ & E & O & E$_{(l=i)}$ & O & O$_{(l=i)}$ & E$_{(l=i)}$\\
& & E$_{(l\not=i)}$ & & & O$_{(l\not=i)}$ & & E$_{(l\not=i)}$ & O$_{(l\not=i)}$\\
${\cal C}$ & E/O & O/E & O/E & E/O & E/O & E/O & O/E & O/E\\
${\cal A}$ & V & I & I & V & I & I & V & V\\       
\hline\hline
\end{tabular}
\caption{Transformation properties of the dimension three nonlocal operators  $O_{\Gamma\pm}(\delta z)$. $i,j,k\not=3$. Other notations are the same as in Table~\ref{TAB:PTC_local_eo}.}
\label{TAB:PTC_even-oddness}
\end{center}
\end{table}

The transformation properties of 
$O_{\Gamma\pm}(\delta z)$ 
are listed in Table~\ref{TAB:PTC_even-oddness}. 
We see that ${\cal P}_{\mu}$, ${\cal T}_{\mu}$ and ${\cal C}$ symmetries cannot protect the mixing between $\bf{1}$ and $\gamma_3$ or between $i\gamma_i\gamma_5$ and $\epsilon_{ijk}\sigma_{jk}$ operators of dimension three. This can be summarized as
\begin{eqnarray}
\label{3p8}
O_{\Gamma\pm}(\delta z)\xrightarrow[]{\text{mixes with}}
(1+G_3(\Gamma))O_{\gamma_3\Gamma\mp}(\delta z),
\end{eqnarray}
which is consistent with the mixing pattern found using lattice perturbation theory in Refs.~\cite{Constantinou:2017sej, Green:2017xeu}. However, if the lattice theory preserves axial or chiral symmetry, then none of the dimension three
operators will mix with each other.

The mixing among dimension three operators of different $\delta z$ cannot be excluded by symmetries but diagrammatic analysis excludes this possibility to all  orders in the strong coupling constant expansion \cite{Ji:2017oey}. The mixing of dimension three to dimension four operators of different $\delta z$ is not systematically studied yet. However, the one loop example in Eq.(\ref{48}) is consistent with no mixing among operators of different $\delta z$.

\subsection{Dimension four nonlocal operators}

Now, we extend the discussion for $p$ type and $m$ type local operators to nonlocal ones. We can insert $\slashed{D}$ at any point on the Wilson line. The symmetry properties will not depend on where $\slashed{D}$ is inserted.

\begin{eqnarray}
Q_{\Gamma\!\overrightarrow{D}_{\alpha}}(\delta z,\delta z')&=&
\overline{\psi}(x+\hat{\bm 3}\delta z)
U_3(x+\hat{\bm 3}\delta z;x+\hat{\bm 3}\delta z')\Gamma\overrightarrow{\slashed{D}}_{\alpha}
U_3(x+\hat{\bm 3}\delta z';x)
\psi(x),
\label{EQ:Opa_non-local_rGD-3}\\
Q_{\overrightarrow{D}_{\alpha}\Gamma}(\delta z,\delta z')&=&
\overline{\psi}(x+\hat{\bm 3}\delta z)
U_3(x+\hat{\bm 3}\delta z;x+\hat{\bm 3}\delta z')\overrightarrow{\slashed{D}}_{\alpha}\Gamma
U_3(x+\hat{\bm 3}\delta z';x)
\psi(x),
\label{EQ:Opa_non-local_rDG-3}\\
Q_{\Gamma\!\overleftarrow{D}_{\alpha}}(\delta z,\delta z')&=&
\overline{\psi}(x+\hat{\bm 3}\delta z)
U_3(x+\hat{\bm 3}\delta z;x+\hat{\bm 3}\delta z')
\Gamma
\overleftarrow{\slashed{D}}_{\alpha}U_3(x+\hat{\bm 3}\delta z';x)\psi(x),
\label{EQ:Opa_non-local_lGD-3}\\
Q_{\overleftarrow{D}_{\alpha}\Gamma}(\delta z,\delta z')&=&
\overline{\psi}(x+\hat{\bm 3}\delta z)U_3(x+\hat{\bm 3}\delta z;x+\hat{\bm 3}\delta z')\overleftarrow{\slashed{D}}_{\alpha}
\Gamma U_3(x+\hat{\bm 3}\delta z';x)\psi(x),
\label{EQ:Opa_non-local_lDG-3}
\end{eqnarray}
where $0\le \delta z' \le  \delta z$.
The $z$-direction is treated differently by writing $\alpha\in[3,\perp]$ and $\overrightarrow{\slashed{D}}_3=\gamma_3\overrightarrow{D}_3$, and $\overrightarrow{\slashed{D}}_{\perp}=\sum_{\mu\not=3}\gamma_{\mu}\overrightarrow{D}_{\mu}$. 

\begin{table}[t]
\begin{center}
\small
\begin{tabular}{ccccccccc}
\hline\hline
& $\Gamma=\bf{1}_{+/-}$ & $\gamma_{i+/-}$ & $\gamma_{3+/-}$ & $\gamma_{5+/-}$ & $\gamma_i\gamma_{5+/-}$ & $\gamma_3\gamma_{5+/-}$ & $\sigma_{i3+/-}$ & $\epsilon_{ijk}\sigma_{jk+/-}$\\
\hline
${\cal P}_3$ & E & O & E & O & E & O & O & E\\       
${\cal P}_{l\not=3}$ & E/O & E/O$_{(l=i)}$ & O/E & O/E & O/E$_{(l=i)}$ & E/O & O/E$_{(l=i)}$ & E/O$_{(l=i)}$\\
& & O/E$_{(l\not=i)}$ & & & E/O$_{(l\not=i)}$ & & E/O$_{(l\not=i)}$ & O/E$_{(l\not=i)}$\\
${\cal T}_3$ & E/O & E/O & O/E & O/E & O/E & E/O & O/E & E/O\\
${\cal T}_{l\not=3}$ & E & O$_{(i=l)}$ & E & O & E$_{(l=i)}$ & O & O$_{(l=i)}$ & E$_{(l=i)}$\\
& & E$_{(l\not=i)}$ & & & O$_{(l\not=i)}$ & & E$_{(l\not=i)}$ & O$_{(l\not=i)}$\\ 
${\cal C}$($Q_{\Gamma\pm/\overline{\Gamma}\pm}^{{D}_{\alpha}(+)}$) & O/E & E/O & E/O & O/E & O/E & O/E & E/O & E/O\\
${\cal C}$($Q_{\Gamma\pm/\overline{\Gamma}\pm}^{{D}_{\alpha}(-)}$) & E/O & O/E & O/E & E/O & E/O & E/O & O/E & O/E\\
$\cal A$ & I & V & V & I & V & V & I & I\\       
\hline\hline
\end{tabular}
\caption{Transformation properties of the dimension four $p$ type
 nonlocal operators  $Q_{\Gamma\pm/\overline{\Gamma}\pm}^{{D}_{\alpha}(\pm)}(\delta z,\delta z')$. $i,j,k\not=3$. Other notations are the same as in Table~\ref{TAB:PTC_local_eo}.
}
\label{TAB:PTC_Opa_even-oddness}
\end{center}
\end{table}

As in the local quark bilinear case, inserting $\overrightarrow{\slashed{D}}$ and $\overleftarrow{\slashed{D}}$ does not change the transformation properties under ${\cal P}_{\mu}$ and ${\cal T}_{\mu}$. These operators transform in the same way as $O_{\Gamma}(\delta z)$. It is useful to define combinations that are even or odd under ${\cal P}_{\mu}$ and ${\cal T}_{\mu}$:
\begin{eqnarray}
Q_{\Gamma\!\overrightarrow{D}_{\alpha}\pm/\overrightarrow{D}_{\alpha}\Gamma\pm}(\delta z,\delta z')
&=&
\frac{1}{2}\left(
Q_{\Gamma\!\overrightarrow{D}_{\alpha}/\overrightarrow{D}_{\alpha}\Gamma}(\delta z,\delta z')\pm
Q_{\Gamma\!\overrightarrow{D}_{\alpha}/\overrightarrow{D}_{\alpha}\Gamma}(-\delta z,-\delta z')
\right),
\\
Q_{\Gamma\!\overleftarrow{D}_{\alpha}\pm/\overleftarrow{D}_{\alpha}\Gamma\pm}(\delta z,\delta z')
&=&
\frac{1}{2}\left(
Q_{\Gamma\!\overleftarrow{D}_{\alpha}/\overleftarrow{D}_{\alpha}\Gamma}
(\delta z,\delta z')\pm
Q_{\Gamma\!\overleftarrow{D}_{\alpha}/\overleftarrow{D}_{\alpha}\Gamma}
(-\delta z,-\delta z')
\right).
\end{eqnarray}
Under $\cal C$, those operators transform as
\begin{eqnarray}
Q_{\Gamma\!\overrightarrow{D}_{\alpha}\pm/\overrightarrow{D}_{\alpha}\Gamma\pm}(\delta z,\delta z')
&\xrightarrow[]{\cal C}&
\mp Q_{\overleftarrow{D}_{\alpha}(C\Gamma C^{-1})^{\top}\pm
/(C\Gamma C^{-1})^{\top}\!\overleftarrow{D}_{\alpha}\pm}(\delta z,\delta z'),
\\
Q_{\overleftarrow{D}_{\alpha}\Gamma\pm/\Gamma\!\overleftarrow{D}_{\alpha}\pm}(\delta z,\delta z')
&\xrightarrow[]{\cal C}&
\mp O_{(C\Gamma C^{-1})^{\top}\!\overrightarrow{D}_{\alpha}\pm
/\overrightarrow{D}_{\alpha}(C\Gamma C^{-1})^{\top}\pm}(\delta z,\delta z').
\end{eqnarray}
So we define the combinations
\begin{eqnarray}
Q_{\Gamma\pm/\overline{\Gamma}\pm}^{{D}_{\alpha}(+)}(\delta z,\delta z')&=&
Q_{\overleftarrow{D}_{\alpha}\Gamma\pm/\Gamma\!\overleftarrow{D}_{\alpha}\pm}(\delta z,\delta z')
+Q_{\Gamma\overrightarrow{D}_{\alpha}\pm/\overrightarrow{D}_{\alpha}\!\Gamma\pm}(\delta z,\delta z'),
\label{3p17}
\\
Q_{\Gamma\pm/\overline{\Gamma}\pm}^{{D}_{\alpha}(-)}(\delta z,\delta z')&=&
Q_{\overleftarrow{D}_{\alpha}\Gamma\pm/\Gamma\!\overleftarrow{D}_{\alpha}\pm}(\delta z,\delta z')
-Q_{\Gamma\overrightarrow{D}_{\alpha}\pm/\overrightarrow{D}_{\alpha}\!\Gamma\pm}(\delta z,\delta z'),
\label{3p18}
\end{eqnarray}
such that
\begin{eqnarray}
Q_{\Gamma\pm/\overline{\Gamma}\pm}^{{D}_{\alpha}(+)}(\delta z,\delta z')
&\xrightarrow[]{\cal C}&
\mp Q_{(C\Gamma C^{-1})^{\top}\pm/(C\overline{\Gamma}C^{-1})^{\top}\pm}^{{D}_{\alpha}(+)}(\delta z,\delta z'),
\\
Q_{\Gamma\pm/\overline{\Gamma}\pm}^{{D}_{\alpha}(-)}(\delta z,\delta z')
&\xrightarrow[]{\cal C}&
\pm Q_{(C\Gamma C^{-1})^{\top}\pm/(C\overline{\Gamma}C^{-1})^{\top}\pm}^{{D}_{\alpha}(-)}(\delta z,\delta z').
\end{eqnarray}
Their properties under ${\cal P}_{\mu}$, ${\cal T}_{\mu}$ and ${\cal C}$ are listed in Table~\ref{TAB:PTC_Opa_even-oddness}. By comparing with Table~\ref{TAB:PTC_even-oddness}, we find that ${\cal P}_{\mu}$, ${\cal T}_{\mu}$ and ${\cal C}$ symmetries do not protect $O_{\Gamma}(\delta z)$ from mixing with $Q_{\Gamma}^{{D}_{\alpha}}(\delta z,\delta z')$ or $Q_{\gamma_3\Gamma}^{{D}_{\alpha}}(\delta z,\delta z')$. If the lattice theory preserves axial or chiral symmetry, then the mixing with $Q_{\Gamma}^{{D}_{\alpha}}(\delta z,\delta z')$ is forbidden, but the mixing with $Q_{\gamma_3\Gamma}^{{D}_{\alpha}}(\delta z,\delta z')$ is still allowed. Since the Wilson line can be described as a heavy-quark propagator in the auxiliary-field approach~\cite{Ji:2015jwa, Ji:2017oey, Green:2017xeu}, this is analogous to the static heavy-light system, which has $p$ type discretization errors even if the light quarks respect chiral symmetry. 
Note that Ref.~\cite{Ishikawa:2018wun} missed the operators with $\delta z'$ different from $0$ and $\delta z$. Since there are many more $p$ type operators now, it makes the non-perturbative improvement program advocated in Ref.~\cite{Ishikawa:2018wun} much more difficult and perhaps unpractical.

The $m$ type nonlocal bilinear is 
\begin{eqnarray}
\label{3p21}
Q_{\Gamma}^{\rm M}(\delta z)=
m\overline{\psi}(x+\hat{\bf 3}\delta z)\Gamma
U_3(x+\hat{\bf 3}\delta z;x)\psi(x).
\end{eqnarray}
It has the same transformation properties as $O_{\Gamma}(\delta z)$ under ${\cal P}_{\mu}$, ${\cal T}_{\mu}$ and ${\cal C}$ but is different for the chiral rotation. However, chiral symmetry does not prevent 
$O_{\Gamma}(\delta z)$ from mixing with the $m$ type operator $Q_{\gamma_3\Gamma}^{\rm M}(\delta z)$.

\subsection{A mixing example in perturbative theory} 
\label{sec3p3}

\begin{figure}
\centering
\includegraphics[scale=0.32, viewport = 0 0 340 260, clip]
{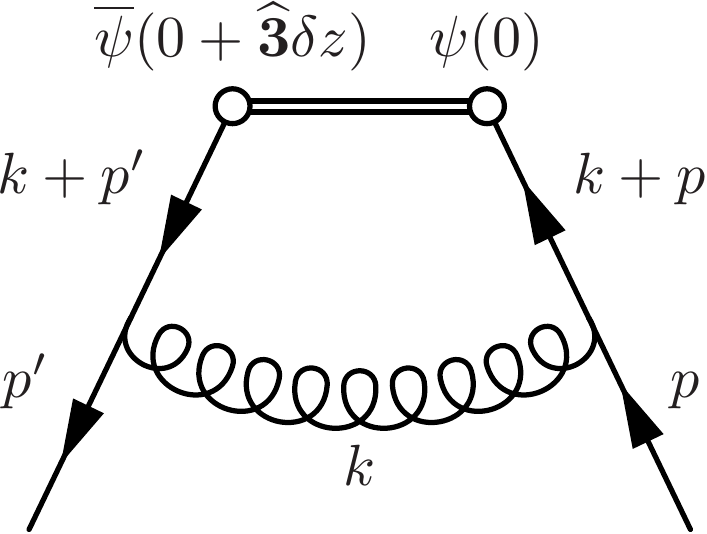}
\caption{One of the one-loop Feynman diagrams for the nonlocal quark bilinear. $p$ and $p'$ are incoming and outgoing external momenta, respectively.}
\label{FIG:vertex-type}
\end{figure}

In the previous section, it was shown that ${\cal P}_{\mu}$, ${\cal T}_{\mu}$, ${\cal C}$, and chiral symmetries cannot protect dimension three 
nonlocal quark bilinears from mixing with dimension four operators. This is a distinct feature from local quark bilinears in which dimension three operators are protected from mixing with dimension four operators. Here, we use the diagram shown in Fig.~\ref{FIG:vertex-type} to demonstrate where the effect comes from. For our purpose, we can simplify our calculation by taking the Feynman gauge and the limit of small external momenta and quark masses, and we will work in the continuum limit with appropriate UV and IR regulators imposed implicitly. Then, the one-loop amputated Green function in figure~\ref{FIG:vertex-type}, $\Lambda_{\Gamma, \delta z}^\text{1-loop}(p', p, m)$, yields

\begin{eqnarray}
\label{48}
&&
\Lambda_{\Gamma, \delta z}^\text{1-loop}(p', p, m) e^{ip_3'\delta z}
\nonumber\\
&=&
\Gamma
+\int_k\frac{\delta_{\mu\nu}\delta_{AB}}{k^2}(-ig\gamma_{\mu}T^A)
\frac{1}{i(\slashed{k}+\slashed{p}')+m}
\Gamma e^{-ik_3\delta z}\frac{1}{i(\slashed{k}+\slashed{p})+m}
(-ig\gamma_{\nu}T^B)
\nonumber\\
&=&
\left(1+g^2G_F{\cal A}_{\Gamma, \delta z}^{}\right)
\Gamma
+g^2G_F{\cal A}_{\Gamma, \delta z}^{m}(1+G_3(\Gamma))m\gamma_3\Gamma
\nonumber\\
&&
+g^2G_F{\cal A}_{\Gamma, \delta z}^{p_3}
i\left\{
(1+G_3(\Gamma))(-\slashed{p}_3'\gamma_3\Gamma-\gamma_3\Gamma\slashed{p}_3)
+(1-G_3(\Gamma))(-\slashed{p}_3'\gamma_3\Gamma+\gamma_3\Gamma\slashed{p}_3)
\right\}
\nonumber\\
&&
+g^2G_F{\cal A}_{\Gamma, \delta z}^{p_{\perp}}
i\left\{
(1+G_3(\Gamma))
(-\slashed{p}_{\perp}'\gamma_3\Gamma-\gamma_3\Gamma\slashed{p}_{\perp})
+(1-G_3(\Gamma))
(-\slashed{p}_{\perp}'\gamma_3\Gamma+\gamma_3\Gamma\slashed{p}_{\perp})
\right\}
\nonumber\\
&&
+g^2G_F{\cal A}_{\overline{\Gamma}, \delta z}^{p_{\perp}}
i\left\{
(1+G_3(\Gamma))
(-\gamma_3\Gamma\slashed{p}_{\perp}'-\slashed{p}_{\perp}\gamma_3\Gamma)
+(1-G_3(\Gamma))
(-\gamma_3\Gamma\slashed{p}_{\perp}'+\slashed{p}_{\perp}\gamma_3\Gamma)
\right\}
\nonumber\\
&&+{\cal O}(p'^2, p^2, p'p, p'm, pm, m^2),
\end{eqnarray}
where the coefficients are
\begin{eqnarray}
{\cal A}_{\Gamma, \delta z}^{}
&=&
\int_k\frac{\cos(k_3\delta z)}{(k^2)^3}
\frac{H(\Gamma)}{3}\left(
\left(H(\Gamma)-G_3(\Gamma)\right)k^2
+\left(-H(\Gamma)+4G_3(\Gamma)\right)k_3^2
\right),
\\
{\cal A}_{\Gamma, \delta z}^{m}
&=&
\int_k\frac{\sin(k_3\delta z)k_3}{(k^2)^3}
\left(-H(\Gamma)+2G_3(\Gamma)\right),
\\
{\cal A}_{\Gamma, \delta z}^{p_3}
&=&
\int_k\frac{\sin(k_3\delta z)k_3}{(k^2)^4}
\frac{H(\Gamma)}{6}\left(\left(-2H(\Gamma)+5G_3(\Gamma)\right)k^2
+2\left(H(\Gamma)-4G_3(\Gamma)\right)k_3^2\right),
\\
{\cal A}_{\Gamma, \delta z}^{p_{\perp}}
&=&
\int_k\frac{\sin(k_3\delta z)k_3}{(k^2)^4}
\frac{G_3(\Gamma)}{6}\left(\left(H(\Gamma)-6G_3(\Gamma)\right)k^2
+2H(\Gamma)k_3^2\right),
\\
{\cal A}_{\overline{\Gamma}, \delta z}^{p_{\perp}}
&=&
\int_k\frac{\sin(k_3\delta z)k_3}{(k^2)^4}
\frac{1}{3}\left(\left(-H(\Gamma)+3G_3(\Gamma)\right)k^2+H(\Gamma)k_3^2\right),
\end{eqnarray}
and where $H(\Gamma)=\sum_{\mu=1}^4G_{\mu}(\Gamma)$. It is easy to see that when $\delta z = 0$ (corresponding to a local quark bilinear), the mixings with all dimension four operators vanish, but when $\delta z \ne 0$ (corresponding to a nonlocal quark bilinear), the mixing with dimension four operators appears even though the theory has ${\cal P}$, ${\cal T}$, ${\cal C}$, and chiral symmetries.


\section{Summary}
\label{SEC:summary}

We have used the symmetry properties of nonlocal quark bilinear operators under parity, time reversal and chiral or axial transformations to study the possible mixing among these operators. Below, we summarize our findings.  

\begin{enumerate}
\item If the lattice theory preserves chiral symmetry, then the dimension three nonlocal quark bilinear operators $O_{\Gamma\pm}(\delta z)$ of Eq.~\eqref{3p5} are protected from mixing with each other, but they are not protected from mixing with the dimension four operators of Eqs.~\eqref{3p17}, \eqref{3p18} and \eqref{3p21} with all possible values of $\delta z'$ satisfying $0\le \delta z' \le  \delta z$:  
\begin{eqnarray}
O_{\Gamma}^{p}(\delta z,\delta z') 
&=&(1+G_3(\Gamma))Q_{\gamma_3\Gamma}^{{D}_{\alpha}(-)}(\delta z,\delta z')
+(1-G_3(\Gamma))O_{\gamma_3\Gamma}^{{D}_{\alpha}(+)}(\delta z,\delta z'),
\label{EQ:nonlocal-Opa1}
\\
O_{\overline{\Gamma}}^{p}(\delta z,\delta z') 
&=&(1+G_3(\Gamma))
Q_{\overline{\gamma_3\Gamma}}^{{D}_{\alpha}(-)}(\delta z,\delta z')
+(1-G_3(\Gamma))
O_{\overline{\gamma_3\Gamma}}^{{D}_{\alpha}(+)}(\delta z,\delta z'),
\label{EQ:nonlocal-Opa2}
\\
O_{\Gamma}^{m}(\delta z) 
&=&(1+G_3(\Gamma))Q_{\gamma_3\Gamma}^{\rm M}(\delta z),
\label{EQ:nonlocal-Oma}
\end{eqnarray}
where $G_{\mu}$ is defined in Eq.~\eqref{Gmu}. This mixing pattern is confirmed by an example calculation for a one-loop diagram, as shown in Sec.~\ref{sec3p3}.  \Blue{Since there are many operators in Eqs.~\eqref{EQ:nonlocal-Opa1}--\eqref{EQ:nonlocal-Oma}, it is unpractical to remove the mixing using the improvement procedure.}

\item If the lattice theory breaks chiral symmetry, then the dimension three nonlocal quark bilinear $O_{\Gamma\pm}(\delta z)$ 
mixes with
\begin{eqnarray}
(1+G_3(\Gamma))O_{\gamma_3\Gamma\pm}(\delta z) .
\end{eqnarray}
The operator $O_{\Gamma\pm}(\delta z)$ not only mixes with all the operators in Eqs.~\eqref{EQ:nonlocal-Opa1}--\eqref{EQ:nonlocal-Oma}, but also with $Q_{\Gamma}^{{D}_{\alpha}(-)}(\delta z,\delta z')$, $Q_{\overline{\Gamma}}^{{D}_{\alpha}(-)}(\delta z,\delta z')$ and $Q_{\Gamma}^{\rm M}(\delta z)$ for all possible values of $\delta z'$ satisfying $0\le \delta z' \le  \delta z$.
\end{enumerate}

This study is particularly relevant for the quasi-PDF approach, which receives power corrections in inverse powers of hadron momentum. It is important to find a window where hadron momentum is large enough to suppress power corrections, but at the same time the mixing with $p$ type dimension four operators is under control. For future work, in light of the similarity between the Wilson line and the heavy-quark propagator, it would be valuable to apply techniques developed for heavy-quark effective field theory on the lattice~\cite{Sommer:2010ic,Sommer:2015hea} and the associated treatments to improve lattice artifacts~\cite{Becirevic:2003hd,Blossier:2007hg,Ishikawa:2011dd,Sheikholeslami:1985ij,Heatlie:1990kg,Borrelli:1992vf}.



\acknowledgments

JWC is partly supported by the Ministry of Science and Technology, Taiwan, under Grant No. 105-2112-M-002-017-MY3 and the Kenda Foundation. TI is supported by Science and Technology Commission of Shanghai Municipality (Grants No. 16DZ2260200). TI and LCJ are supported by the Department of Energy, Laboratory Directed Research and Development (LDRD) funding of BNL, under contract DE-EC0012704. The work of HL is supported by US National Science Foundation under grant PHY 1653405. JHZ is supported by the SFB/TRR-55 grant ``Hadron Physics from Lattice QCD'', and a grant from National Science Foundation of China (No.~11405104). YZ is supported by the U.S. Department of Energy, Office of Science, Office of Nuclear Physics, from DE-SC0011090 and within the framework of the TMD Topical Collaboration.

\bibliographystyle{apsrev}
\bibliography{opmx}

\end{document}